\documentclass{amsproc}

\usepackage{amssymb}
\usepackage{amsmath}
\usepackage{amsbsy}
\usepackage{amsfonts}
\usepackage{amsthm}
\usepackage{latexsym}
\usepackage{mathrsfs}
\usepackage{graphicx}
\usepackage{verbatim}
\usepackage{hyperref}
\usepackage{url}

\usepackage{tikz}
\usetikzlibrary{decorations.markings,arrows,backgrounds,patterns}

\newcommand{\beq}{\begin{equation}}
\newcommand{\eeq}{\end{equation}}
\newcommand{\bea}{\begin{eqnarray}}
\newcommand{\eea}{\end{eqnarray}}

\def\Tr{\mathop{\rm Tr}}

\newcommand{\be}{\begin{equation}}
\newcommand{\ee}{\end{equation}}

\newcommand{\m}{\mu}
\newcommand{\n}{\nu}
\newcommand{\s}{\sigma}

\newcommand{\ve}{\epsilon}

\newcommand{\chir}{\bar}

\newcommand{\Id}{1 \hspace{-2.7pt} \text{l}}
\newcommand{\Q}{\mathcal{Q}}
\newcommand{\N}{\mathcal{N}}
\newcommand{\li}{{(\ell)}}
\newcommand{\copl}{\mathbf{q}}
\newcommand{\q}{{\tt q}}

\newtheorem{thm}{Theorem}

\def\XXint#1#2#3{{\setbox0=\hbox{$#1{#2#3}{\int}$}
     \vcenter{\hbox{$#2#3$}}\kern-.5\wd0}}

\begin{document}


%
%

\begin{flushright}
\normalsize
~~~~
SISSA 34/2016/MATE-FISI
\end{flushright}

\vspace{80pt}


\begin{center}
{\Large \bf Gauge theories on compact toric surfaces, \\
conformal field theories and equivariant Donaldson invariants}
\end{center}

\vspace{25pt}

\begin{center}
{
Mikhail Bershtein$^{\spadesuit}$, Giulio Bonelli$^{\heartsuit}$, Massimiliano Ronzani$^{\heartsuit}$ and Alessandro Tanzini$^{\heartsuit}$
}\\
%
\vspace{15pt}

$^{\spadesuit}$ Landau Institute for Theoretical Physics, Chernogolovka, Russia,\\
National Research University Higher School of Economics, International Laboratory of Representation Theory
and Mathematical Physics,\\
Institute for Information Transmission Problems, Moscow, Russia,\\
Independent University of Moscow, Moscow, Russia \footnote{email: mbersht@gmail.com}\\

\vspace{15pt}
%
$^{\heartsuit}$International School of Advanced Studies (SISSA) \\via Bonomea 265, 34136 Trieste, Italy 
and INFN, Sezione di Trieste \footnote{email: bonelli,mronzani,tanzini@sissa.it}\\

\end{center}
%

\vspace{20pt}

\noindent
\textsc{Abstract:}
We show that equivariant Donaldson polynomials of compact toric surfaces can be calculated as 
residues of suitable combinations of Virasoro conformal blocks, by building on AGT correspondence
between $\mathcal{N}=2$ supersymmetric gauge theories and two-dimensional conformal field theory.

Talk\footnote{\url{http://salafrancesco.altervista.org/wugo2015/tanzini.pdf}}
presented by A.T. at the conference
\emph{Interactions between Geometry and Physics -- in honor of Ugo Bruzzo's 60th birthday}
17-22 August 2015, Guaruj\'a, S\~{a}o Paulo, Brazil,
mostly based on \cite{Bawane:2014uka} and \cite{Bershtein:2015xfa}.





\tableofcontents


\section{Introduction}


A link between $\mathcal{N}=2$ supersymmetric gauge theories in four dimensions 
and Donaldson invariants, classifying differentiable structures on four-manifolds, has been established since the seminal paper by Witten \cite{Witten:1988ze}. 
Here we exploit recent progresses both in the formulation of supersymmetric quantum field theories
on curved spaces and in equivariant localization applied to supersymmetric path integrals in order to provide a direct computation
of {\it equivariant} Donaldson polynomials for compact toric surfaces along the lines suggested by \cite{Nekrasov:2003viNO}.
A crucial new ingredient along this path is provided by the correspondence 
between supersymmetric field theories in four dimensions and two-dimensional conformal field theories \cite{Alday:2009aq}.
Indeed, our final result is that equivariant Donaldson polynomials on a compact toric manifold $X$ can be expressed as residues of suitable combinations
of Virasoro conformal blocks. This follows from the fact that the supersymmetric path integral reduces to a contour integral over
a product of toric patches contributions given by Nekrasov partition functions \cite{Nekrasov:2002qd}. These latter depend on the weights of the toric action
through effective Omega-background parameters and correspond to Virasoro conformal blocks whose central charge depend on these parameters.
The integration variable is the v.e.v.~of the scalar field of the $\mathcal{N}=2, d=4$ vector multiplet, which on compact manifolds is a normalizable
 mode to be integrated over in the space of admissible classical BPS solutions.
The latter are parameterized also by magnetic fluxes labeled by an integral lattice.
A subtle issue concerns the allowed values of these fluxes, reflecting the stability conditions to be imposed on the equivariant vector 
 bundles. We fully solved this problem for $X=\mathbb{P}^2$ in \cite{Bershtein:2015xfa}
 and we present here some recent progress on $X=\mathbb{P}^1\times\mathbb{P}^1$.
 
 We also consider maximally supersymmetric gauge theories and express the resulting partition functions,
 generating the Euler characteristics of the moduli spaces of stable
 equivariant vector bundles, in terms of mock modular forms. 
 It is known from \cite{Vafa:1994tf} that these partition functions, labeled by the first Chern class of the bundle,
 form a non-trivial representation  of the modular group $SL(2,\mathbb{Z})$, obeying an analogue of the Verlinde algebra
 satisfied by the conformal blocks of rational conformal field theories in two dimensions.
 
 On the other hand, it is known that knot invariants can be calculated by suitable gluing of 
Wess-Zumino-Witten conformal blocks \cite{Witten:1988hf}. The result we find here is somewhat analogous in the fact
 that equivariant Donaldson invariants are obtained from the gluing of conformal blocks of a {\it non-rational} conformal field theory.
 It would be interesting to further investigate this analogy  for example by considering the insertion of surface operators in the supersymmetric path integral,
 which are known to be related to $\widehat{sl}(2)$ conformal blocks \cite{Alday:2010vg}.
 Surface operators in four-dimensions provide indeed co-boundary operators for knots \cite{2008arXiv0806.1053K}.

\section{Supersymmetry on compact manifolds}

The symmetry group for the $\N=2$ supersymmetric theory on $\mathbb{R}^4$ is
given by the rotation group $SO(4)=SU(2)_1\times SU(2)_2$ and the
$R$-symmetry group $SU(2)_R\times U(1)_R$.

The supersymmetry is generated by the operator
$\Q=\xi^{A\alpha}Q_{A\alpha}+\bar\xi^{A}_{\dot\alpha}\bar Q_{A}^{\dot\alpha}$
where the generators $\xi, \bar\xi$ are commuting Weyl spinors of $SU(2)_{1,2}$ (indices $\alpha,\dot\alpha$)
respectively and 
moreover both of them transform in the two-dimensional representation of $SU(2)_R$ (index $A$).

On a generic manifold  the covariantly constant condition $D\xi=0$ is too restrictive.

The consistency of the $\N=2$ supersymmetry algebra
on a compact four manifold requires \cite{Bawane:2014uka}
that the spinor parameters have to satisfy the generalized Killing equations
\be
\begin{aligned}
 &D_\m \xi_B + T^{\rho\sigma}\sigma_{\rho\sigma}\sigma_\m\bar\xi_B - \frac{1}{4} \sigma_{\m}\bar\sigma_{\n}D^{\n}\xi_B=0 \\
 &D_\m \bar\xi_B + \bar T^{\rho\sigma}\bar\sigma_{\rho\sigma}\bar\sigma_\m\xi_B - \frac{1}{4} \bar\sigma_{\m}\sigma_{\n}D^{\n}\bar\xi_B=0
\end{aligned}
\ee
and the auxiliary equations
\begin{equation}
\begin{aligned}
\s^\m \chir\s^\n D_\m D_\n \xi_A + 4 D_\lambda T_{\m\n} \s^{\m\n} \s^\lambda \chir\xi_A &= M_1\xi_A, \\
\chir \s^\m \s^\n D_\m D_\n \chir\xi_A + 4 D_\lambda \chir T_{\m\n} \chir\s^{\m\n} \chir\s^\lambda \xi_A &= M_2\chir\xi_A,
\end{aligned}
\end{equation}
where background fields appear: two scalar $M_1,M_2$ a self-dual tensor $T_{\mu\nu}$,
an anti-self-dual tensor $\bar T_{\mu\nu}$ and the connection of the $SU(2)_R$ $R$-symmetry
bundle hidden in the covariant derivatives $D_\mu$.

This is in agreement with the result of \cite{Hama:2012bg, Klare:2013dka}
obtained from supergravity in the spirit of \cite{Dumitrescu:2012ha, Dumitrescu:2012at}.

\subsection{Four-manifolds admitting a $U(1)$-isometry and the equivariant topological twist}

For a general four-manifold it is always possible to solve  the generalized Killing equations
at least for one scalar supercharge ($\bar\xi_{A}^{\dot\alpha}=\delta_{A}^{\dot\alpha}$, $\xi_{A\alpha}=0$),
performing a topological twist \cite{Witten:1988ze}.
The spinors parameters are sections of the  bundles
\be
\xi\in\Gamma\left(S^+\otimes \mathcal{R}\otimes \mathcal{L}_R\right)
\qquad
\bar\xi\in\Gamma\left(S^-\otimes \mathcal{R}^\dagger\otimes \mathcal{L}_R^{-1}\right)
\ee
where $S^\pm$ are the spinor bundles of chirality $\pm$,
$\mathcal{R}$ is the $SU(2)$ $R$-symmetry vector bundle
and
$\mathcal{L}_R$ is the $U(1)$ $R$-symmetry line bundle.
Choosing $\mathcal{L}_R=\mathcal{O}$ to be the trivial line bundle and $\mathcal{R}=S^-$
realizes the topologically twisted theory,
that is matching the $R$-symmetry connection with the spin connection of the manifold.
For this choice of the $R$-symmetry bundles,
$S^+\otimes S^-\sim T$ and $S^-\otimes S^- \sim \mathcal{O}+ T^{(2,+)}$
with $T$ the tangent bundle and $T^{(2,+)}$ the bundle of self-dual two-forms.

If the manifold has a $U(1)$-isometry generated by the vector field $V$
a more general solution is available \cite{Nekrasov:2002qd,Klare:2013dka,Bawane:2014uka}.
\be
\bar\xi_{A}^{\dot\alpha}=\delta_{A}^{\dot\alpha},
\qquad
\xi_{A\alpha}=V^{\mu}(\sigma_{\mu})_{\alpha\dot\alpha}\xi_A^{\dot\alpha},
\qquad
T=-\frac{1}{32}(d\zeta)^{-},
\qquad
\bar T=M_1=M_2=0,
\ee
where $\zeta$ is a $U(1)$-invariant one-form such that $\iota_V \zeta=(V,V)$.

The supercharge generated by this solution has a scalar $Q$ and a vector-like $Q_\mu$ component
\be
\Q=Q+V^{\mu}Q_{\mu}
\ee
and generates the following twisted $\N=2$ superalgebra
\be\label{eq:newSUSY1}
\begin{aligned}
&\Q A=\Psi, &\quad &\Q\Psi=i\iota_V F + D\Phi, &\quad& \Q \Phi= i \iota_V\Psi,\\
&\Q\chir\Phi=\eta, &\quad &\Q\eta=i\,\iota_VD\chir\Phi+i[\Phi,\chir\Phi], \\
&\Q\chi^+=B^+, &\quad &\Q B^+=i\mathcal{L}_V\chi^++i[\Phi,\chi^+]
\end{aligned}
\ee
on twisted fields: a connection $A_\mu$ and two complex scalars $\Phi, \bar \Phi$ with even statistics,
and a scalar $\eta$, a one-form $\Psi_\mu$ and a self-dual two-form $\chi^{+}$ with odd statistics.
The supercharge $\Q$  acts as a differential on the fields and squares to bosonic symmetries
$\Q^2=i \mathcal{L}_V +\delta^\text{gauge}_\Phi$.

\section{Observables and equivariant Donaldson polynomials}
\label{section-obs}

Equivariant observables in the topologically twisted theory
are obtained by the equivariant version of the usual descent equations.

The supersymmetry action can be written as the equivariant Bianchi identity for the 
curvature ${\bf F}$ of the universal bundle as \cite{Baulieu:2005bs}
\be
{\bf D F}\equiv
\left(-Q + D +i\iota_V\right)\left(F+\psi+\Phi\right)=0,
\ee
where $D$ is the covariant derivative.
Observables $\mathcal O$ are obtained by the insertion of ad-invariant polynomial  ${\mathcal P}$
 on the Lie algebra of the gauge group,
 intersected with elements of the equivariant cohomology
of the manifold, ${\bf \Omega}\in H^\bullet_V(X)$
\be
{\mathcal O}\left({\bf \Omega},{\mathcal P}\right)\equiv
\int {\bf \Omega}\wedge {\mathcal P}({\bf F}), 
\qquad
Q{\mathcal P}({\bf F})=\left(d+i\iota_V\right){\mathcal P}({\bf F}).
\ee

The relevant observables
in the $U(2)$ theory are obtained from ${\mathcal P}_n(x)=\frac{1}{n}\Tr x^n$ with $n=1,2$.
The first,  $\int_X \Tr {\bf F} \wedge \Omega$,
is the source term for the first Chern class and for the local observable $\Tr\Phi(P)$, where
$P$ is a fixed point of the vector field $V$.
The second, $\frac{1}{2}\int_X  \Tr {\bf F}^2 \wedge \Omega^{[\text{even}]}$,
generates
\begin{itemize}
\item the second Chern character of the gauge bundle $\int_X \Tr (F\wedge F)$ for ${\bf \Omega}=1$ 
      (the Poincar\'e dual of $X$),
\item surface observables for ${\bf \Omega}=\omega+H$, where $\omega$ is a V-equivariant element 
      in $H^2(X)$ and $H$ a linear polynomial in the weights of the V-action
      satisfying $d H=\iota_V\omega$.
      Namely
      \be\label{16}
      \int_X \omega\wedge \Tr\left(\Phi F +\Psi^2\right) + H \Tr (F\wedge F)
      \ee
\item local observables at the fixed points: $\Tr\Phi^2(P)$, for ${\bf \Omega}=\delta_P$ 
      the Poincar\'e dual of any fixed point $P$ under the $V$-action.
\end{itemize}

The local observables in the equivariant case depend on the 
insertion point via the equivariant weights of the fixed point.
This is due to the fact that the equivariant classes of different fixed points are distinct.
From the gauge theory viewpoint one has
\be
\Tr\Phi^2(P)-\Tr\Phi^2(P')=\int_{P'}^P \iota_V \Tr \left(\Phi F +\frac{1}{2}\Psi^2\right) + Q[\ldots]
\ee
so that the standard argument of point location independence is flawed by the first 
term in the r.h.s..


The mathematical meaning of these facts is that the equivariant Donaldson polynomials
give a finer characterization of differentiable manifolds. 
The physical one is that the $\Omega$-background probes the gauge theory via a finer 
BPS structure.

\section{$\N=2$ supersymmetric path integral and localisation}

We consider $\N=2$ theory on four dimensional K\"ahler manifold $X$
with gauge group $U(N)$.

In this case the BPS supersymmetric configurations minimizing the path integral are
given by a mild generalization of anti-instantons \cite{Vafa:1994tf}
called Hermitian--Yang--Mills (HYM) connections, satisfying
\be\label{khym}
F^{(2,0)}=0, \qquad
\omega\wedge F=\lambda\, \omega\wedge\omega \Id,
\qquad
\text{where}
\qquad
\lambda=
\frac{2\pi\int_X c_1(E)\wedge\omega}
{r(E)\int_X \omega\wedge\omega}
= \frac{2\pi\mu(E)}
{\int_X \omega\wedge\omega}.
\ee
$\mu(E)$ is the {\it slope} of the vector bundle. Here $r(E)=N$ is the rank of $E$ and $c_1(E)=\frac{1}{2\pi}\text{Tr} F_E$
its first Chern class.

The Hitchin--Kobayashi correspondence establishes the equivalence of the
HYM condition with the semi-stability of the bundle,
it was proven in \cite{Donaldson, Uhlenbeck1, Uhlenbeck2}.
We will see that in the evaluation of the path integral it is essential to implement 
stability conditions of the bundles.

The supersymmetric Lagrangian considered is
\be\label{lagrangian}
L=2\pi i \tau \Tr F^{-}\wedge F^{-}+2\pi i \bar\tau \Tr F^{+}\wedge F^{+}
+ \gamma\wedge\Tr F +\Q \mathcal{V}
\ee
where $\tau=\frac{\theta}{2\pi}-\frac{4\pi i}{g^2}$ is the complexified coupling constant,
$\gamma\in H^2(X)$ is the source for the first Chern class of the vector bundle
and $\mathcal{V}$ is a gauge invariant localizing term, chosen in order to 
implement the HYM condition
\be\label{GFFform}
\mathcal{V}=-\text{Tr}\big[
i\chi^{(0,2)}\wedge F^{(2,0)}+
i\chi\left(\omega\wedge F-\lambda\, \omega\wedge\omega \Id\right)
          +\Psi\wedge\star(\Q\Psi)^\dagger +\eta\wedge\star(\Q\eta)^\dagger\,\big].
\ee

The zeros of the localizing action $\Q\mathcal{V}$, i.e. the supersymmetric fixed points
are given by
\be\label{BPScond1}
\iota_VD\chir\Phi+[\Phi,\chir\Phi]=0, \qquad
i\iota_V F + D\Phi=0,
\ee
and their integrability conditions 
\be
\iota_V D\Phi=0, \qquad
{\mathcal L}_V F =[F,\Phi].
\label{integrability}
\ee

\subsection{Solution on toric surfaces}
For a compact toric manifold the BPS fixed points are given explicitly
by the data from the toric fan.
\begin{figure}[!h]
\centering
\begin{tikzpicture}[scale=0.6]
\draw [->] (0,0) -- (1.5,0) node[right] {$D_0$};
\draw [->] (0,0) -- (0,1.5) node[above] {$D_1$};
\draw [->] (0,0) -- (-1.5,-1.5) node[left] {$D_2$};

\node at (1,1) {$\sigma_0$};
\node at (-1,0.5) {$\sigma_1$};
\node at (0.5,-1) {$\sigma_2$};
\end{tikzpicture}
\caption{Toric fan of $\mathbb{P}^2$.
$\sigma_\ell$ labels the cone of dimension two relative to the $\ell$-th $\mathbb{C}^2$ coordinates patch.}
\end{figure}
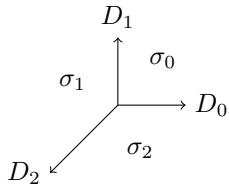

First of all,  using the reality condition on the scalar fields $\chir\Phi=-\Phi^\dagger$, it is possible to see
that $F$ and $\Phi$ are in the same Cartan sub-algebra.
At this point the solution to \eqref{BPScond1} are given by
\be
\left(F+\Phi\right)_\alpha=F^\text{point}_\alpha + 
a_\alpha+\sum_{\ell}k_{\alpha}^\li \omega^\li
\label{phi}
\ee
that is, $F+\Phi$ is the $(\mathbb{C}^\ast)^{N+2}$ equivariant curvature of the bundle.
Moreover  $\omega^\li\in H^2_V(X)$ is the $V$-equivariant two-form Poincar\'e dual
of the equivariant divisor $D_\ell$ corresponding to the $\ell$-th vector of the fan (see figure 1).  
$F^{\rm point}$ is the contribution of point-like instantons located at the fixed points
of the $(\mathbb{C}^\ast)^{2}$-action that
corresponds to ideal sheaves on $T_P X\cong \mathbb{C}^2$ supported
at the fixed points $P$ of the $(\mathbb{C}^\ast)^2$-action,  
these are labeled by Young diagrams $\big\{Y_\alpha^\li\big\}$.

The Cartan algebra valued parameters in \eqref{phi} are
\begin{itemize}
\item $a_\alpha\in \mathbb{C}^N$ are the generators to the $(\mathbb{C}^\ast)^{N}$-action,
they are normalizable constant modes to be eventually integrated over,
\item $k^\li_\alpha\in \mathbb{Z}^{N\chi(X)}$ are gauge field magnetic fluxes subject to stability conditions
imposed by HYM equation.
\end{itemize}

Eventually each of the fixed points of the $(\mathbb{C}^\ast)^{2}$-action on $X$ contributes to the partition function
with a copy of the Nekrasov partition function evaluated on the tangent space of the manifold at the fixed point.

\subsection{Contour of integration and fermionic zero modes}

The treatment of the fermionic zero modes has as a consequence
the identification of the contour of integration to be taken in the evaluation
of the path integral.

The fermionic fields are the scalar $\eta$, the 1-form $\Psi$ and the selfdual 2-form $\chi^+$. 
The number of zero modes is given by the respective Betti numbers
$b_0=1$, $b_1=0$ and $b_2^+=1$ (for all connected and simply connected toric surfaces)
times the rank of the gauge group.
Specifically, the $\chi^+$ zero mode is proportional to the K\"ahler form $\omega$.
We consider the gauge group $U(N)$ with algebra $su(N)\oplus u(1)$.

The zero modes in the $u(1)$ sector come as a quartet of symmetry parameters 
of the whole twisted super-algebra.
The c-number BRST charge implementing this shift symmetry is given by
\be
\begin{aligned}
&\q A=0, &&  \q \Psi= 0, &&  \q\Phi=\kappa_\Phi \Id, && \q\kappa_\Phi=0,\\
&\q \bar\Phi=\kappa_{\bar\Phi}\Id , &&  \q \kappa_{\bar\Phi}= 0, &&  \q\eta=\kappa_\eta \Id, &&  \q\kappa_\eta=0,\\
&\q \chi=\kappa_{\chi}\omega\Id , && \q \kappa_{\chi}= 0, && \q B=0, &&
\end{aligned}
\ee
and the action of $\Q$ on the c-number parameters above is given by
\be
\Q\kappa_\Phi=0, \quad \Q\kappa_{\bar\Phi}=-\kappa_\eta, \quad \Q\kappa_\eta=0, \quad \Q\kappa_\chi=0, 
\ee
so that $\left\{\Q,\q\right\}=0$.
The $\kappa$-ghosts have to be supplemented by their corresponding anti-ghosts
$\bar\kappa_I$ and Lagrange multipliers $\lambda_I$, with $I\in\left\{\Phi,\bar\Phi,\eta,\chi\right\}$
and $\q\bar\kappa_I=\lambda_I$, $\q\lambda_I=0$. 
Moreover $\Q\bar\kappa_I=0$, $\Q\lambda_I=0$ and  $\q\mathcal{V}=0$.
The gauge fixing fermion for the $u(1)$ zero modes then reads
\be
\nu=\sum_I \bar\kappa_I\int_X \Tr (I) e^\omega
\ee
so that the gauge fixing action $(\Q+\q)\nu$ gives a suitable measure to integrate out these modes 
as a perfect quartet.

From the treatment of the zero modes in the $su(N)$ sector 
we get precise instructions 
about the integration on the leftover $N-1$ Cartan parameters $a_\rho=a_\alpha-a_\beta$.
The presence of these zero modes makes the path-integral measure ill-defined,
in order to solve this problem we add to the localizing action the further term
depending only on the $su(N)$ zero modes
\be
S_\text{gauginos}=s\Q \int_{X} \Tr \bar\Phi_0 \chi_0 \omega 
=s\int_{X}\Tr \left\{\eta_0 \chi_0 \omega  
+ \bar\Phi_0 b_0 \omega
\right\}\, .
\label{ga}
\ee
where $s$ is a complex parameter.
The final result does not depend on the actual value of $s$ as long as $s \not= 0$.

Once the integral over the $N-1$ pairs of gluino zero modes $(\eta_0,\chi_0)$ is taken, we stay with 
an insertion of b-field zero mode per $su(N)$ Cartan element as
\be
\prod_\rho\left( \int da\, d\bar a\, db_0\, (s\omega)\, e^{s \bar a b_0\omega}\right)_{\!\!\rho} e^{\Q\mathcal{V}}
\ee
where $\rho$ spans the $su(N)$ Cartan subalgebra.
By renaming $\bar a\to \bar a/s$ and letting $s\to \infty$ we then get 
\be
\prod_\rho\left( \int da\, d\bar a
\frac{\partial}{\partial\bar a}\int \frac{db_0}{b_0}\, e^{\bar a b_0\omega}
\right)_{\!\!\rho} e^{\Q\mathcal{V}|_{\bar a=0}}
\, .
\label{pd}
\ee
Similar arguments appeared in the evaluation of the low-energy effective Seiberg--Witten theory \cite{Losev:1997tp}.
The integrals over the $N-1$ zero modes of $b$ are taken by evaluating at $b=0$ by Cauchy theorem.
This implies that the leftover integral over the Cartan parameters is a total differential in the $\bar \Phi$
zero-mode variables, namely in $\bar a_{\rho}$, so that it gets reduced to a contour integral along the boundary 
of the moduli space of BPS minima studied in the previous subsection.

\subsection{Partition function on compact toric surfaces}

From the analysis of the first subsections we can write the partition function
of the $\mathcal{N}=2$ gauge theory on a compact toric surface $X$ in
a contour integral representation
\be\label{Zfull-gen}
Z^{X}_{\text{full}}\big(\copl,y\,;\ve_1,\ve_2\big)
   =\sum_{\{k^\li_\alpha\}|\text{semi-stable}} \oint d{\vec a}\,
   \prod_{\ell=0}^{\chi(X)} Z_\text{full}^{\mathbb{C}^2}\big(\copl\,;{\vec a}^\li,\ve_1^\li,\ve_2^\li\big)
   \, y^{c_1^{(\ell)}}
\ee
here $\copl=\exp(2\pi i \tau)$ is the exponential of the gauge coupling,
$y$ is the source term corresponding to the K\"ahler form $t \omega$ with $t$ the complexified K\"ahler parameter,
so that $y=e^{2\pi t}$, and $c_1^{(\ell)}=\text{Tr}(\vec k^{(\ell)})$.

The integral is a product of copies of the Nekrasov partition function \cite{Nekrasov:2002qd},
one copy for each fixed point of the $(\mathbb{C}^\ast)^2$ torus action on the manifold $X$, they are 
as many as the Euler number $\chi(X)$.
Moreover $\ve_1^\li,\ve_2^\li$ are the generators of $(\mathbb{C}^\ast)^2$
action on $T_{P_\li}X$ and 
$\vec a^\li$ are the generators of the $(\mathbb{C}^\ast)^N$ framing action,
that are the v.e.v.'s of the scalar field $\Phi$ at the BPS minima \eqref{phi}.

\section{Results for $U(2)$ $\N=2$ theory on the complex projective plane}
\label{section-p2}
We specify formula \eqref{Zfull-gen} for the $U(2)$ gauge theory on $\mathbb{P}^2$,
described by homogeneous coordinates $[z_0:z_1:z_2]$.

The full partition function is therefore given by 
a contour integral formula around the origin
\be\label{Zfull}
Z^{\mathbb{P}^2}_{\text{full}}\big(\copl,x,z,y\,;\ve_1,\ve_2\big)
   =\sum_{\{k^\li_\alpha\}|\text{semi-stable}} \oint d{a}\,
   \prod_{\ell=0}^2 Z_\text{full}^{\mathbb{C}^2}\big(\copl^\li\,;{a}^\li,\ve_1^\li,\ve_2^\li\big)
   \, y^{c_1^{(\ell)}}
\ee
where the generators $\ve_1^\li,\ve_2^\li$ of the $(\mathbb{C}^\ast)^2$
action on $T_{P_\li}\mathbb{P}^2$  are 
\be\label{weights}
[z_0:z_1:z_2]\to[z_0:e^{\ve_1}z_1:e^{\ve_2}z_2] \quad \iff \quad
\begin{array}{c|l|l}
\ell & \ve_1^\li      & \ve_2^\li     \\ \hline
0   & \ve_1           & \ve_2         \\
1   & \ve_2-\ve_1     & -\ve_1        \\
2   & -\ve_2          & \ve_1-\ve_2           
\end{array}
\ee
and the generators $a_\alpha^\li$ of the $(\mathbb{C}^\ast)^2$ framing action are
\be\label{a-patches}
\begin{aligned}
\vec a^{(0)}&=\vec a+\vec p\ve_1+\vec q\ve_2, \\
\vec a^{(1)}&=\vec a+\vec q(\ve_2-\ve_1)+\vec r(-\ve_1), \\
\vec a^{(2)}&=\vec a+\vec p(\ve_1-\ve_2)+\vec r(-\ve_2),
\end{aligned}
\ee
where $\vec p,\vec q,\vec r$ are new names for the magnetic fluxes $\vec k^\li$.
Moreover $\copl^\li=\copl\, e^{\imath^*_{P_{(\ell)}}(\alpha z + p x)}$
is the gauge coupling shifted by the insertion of the observables described in section \ref{section-obs}
evaluated at the $(\mathbb{C}^\ast)^2$-fixed points $P_{(\ell)}$ of $\mathbb{P}^2$.
These insertions are needed for the computation of equivariant Donaldson polynomials.

\subsection{Stability conditions}
The sum over $\vec k^\li\in \mathbb{Z}^{6}$ in \eqref{Zfull} is constrained by the
stability conditions imposed by the HYM equation.
These latter are identified by comparing the data of
the fixed point solutions with Klyachko's description of equivariant
(semi-)stable vector bundles \cite{Klyachko,Kool}.
Defining $p=p_1-p_2$, $q=q_1-q_2$, $r=r_1-r_2$ these are given by
\be\label{stability-p2}
p,q,r\in \mathbb{Z}_{\ge 0}, \quad
p+q\ge r, \quad p+r\ge q, \quad q+r\ge p.
\ee
If we consider the first Chern class $c_1=\text{odd}$, only strictly stable bundles contribute to the path integral
and in \eqref{stability-p2} one can take strict inequalities. In this case the moduli space is smooth.

The case with $c_1=\text{even}$ is more subtle. The contributions saturating one inequality in  \eqref{stability-p2}
correspond to strict semi-stable bundles associated with reducible connections.
These latter are associated with singularities in the moduli space and
to the knowledge of the authors there are no calculations of equivariant Donaldson polynomials in this case.

\subsection{AGT and Zamolodchikov's recursion relations}

\begin{figure}[h!]
\centering
\begin{tikzpicture}[scale=0.5]

\fill[black!20] (1,1) -- (2,1) -- (2,2) -- (1,2);

\draw [-] (0,0) -- (6,0);
\draw [-] (0,1) -- (6,1);
\draw [-] (0,2) -- (5,2);
\draw [-] (0,3) -- (4,3);
\draw [-] (0,4) -- (2,4);

\draw [-] (0,0) -- (0,4);
\draw [-] (1,0) -- (1,4);
\draw [-] (2,0) -- (2,4);
\draw [-] (3,0) -- (3,3);
\draw [-] (4,0) -- (4,3);
\draw [-] (5,0) -- (5,2);
\draw [-] (6,0) -- (6,1);

\node at (1.5,1.5) {$s$};

\draw [<->, thick] (2,-0.5) -- (5,-0.5);
\draw [<->, thick] (-0.5,2) -- (-0.5,4);

\node at (3.5,-1) {$l_Y(s)$};
\node at (-1.5,3) {$a_Y(s)$};
\end{tikzpicture}
\caption{Young diagram.}\label{YD}
\end{figure}
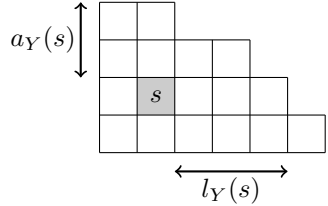

The Nekrasov partition function \cite{Nekrasov:2002qd} appearing in \eqref{Zfull-gen}
can be factorized into three contributions
\be\label{ZfullFact}
	Z^{\mathbb{C}^2}_{\text{full}}(\copl\,;{a},\ve_1,\ve_2)=
        Z^{\mathbb{C}^2}_{\text{class}}(\copl\,;{a},\ve_1,\ve_2)
        Z^{\mathbb{C}^2}_{\text{one-loop}}({a},\ve_1,\ve_2 )
        Z^{\mathbb{C}^2}_{\text{inst}}(\copl\,;{a},\ve_1,\ve_2) 
\ee
with
\begin{align}
 Z^{\mathbb{C}^2}_{\text{class}}(\copl\,;{a},\ve_1,\ve_2)
 &=\exp{\bigg(-\pi i\tau\frac{\sum_\alpha a_\alpha^2}{\ve_1\ve_2}\bigg)} \label{Z-cla} \\
 Z^{\mathbb{C}^2}_{\text{one-loop}}({a},\ve_1,\ve_2 )
 &=\exp{\bigg(-\sum_{\alpha\neq\beta}\gamma_{\ve_1,\ve_2}(a_\alpha-a_\beta)\bigg)} \label{Z-1l} \\
 Z^{\mathbb{C}^2}_{\text{inst}}(\copl\,;{a},\ve_1,\ve_2)
 &=\sum_k \copl^k\, z_k({a},\ve_1,\ve_2) \label{Z-inst}
\end{align}
where the function $\gamma_{\ve_1,\ve_2}$ is the logarithm of
the Barnes double gamma function \cite{Barnes265},
defined by
\be\label{gamma2}
\gamma_{\ve_1,\ve_2}(x)=\frac{d}{d s}\Big|_{s=0}\frac{1}{\Gamma(s)}
                     \int_0^\infty dt\,t^{s-1}\frac{e^{-tx}}{(1-e^{\ve_1 t})(1-e^{\ve_2 t})} \, ,
\ee
and the function $z_k$ in \eqref{Z-inst} is defined
by collections of Young diagrams $\{Y_\alpha\}$ (figure \ref{YD})
via
\be
z_k=\sum_{|\vec Y|=k}\prod_{\alpha,\beta}\prod_{s\in Y_\alpha}
\frac{1}{E(s)(E(s)-\ve_1-\ve_2)}, \qquad
E(s)=a_\alpha-a_\beta-\ve_1\,l_Y(s)+\ve_2(a_Y(s)+1).
\ee

To perform the integral in \eqref{Zfull} it is important to
know the analytic structure of the components in \eqref{ZfullFact}.
It is not difficult to see that the product of the three copies of
$Z^{\mathbb{C}^2}_{\text{one-loop}}({a^\li},\ve^\li_1,\ve^\li_2 )$
has a double zero in the origin if the (semi-)stability conditions \eqref{stability-p2} are satisfied
\be\label{1loop-2zero}
Z^{\mathbb{P}^2}_{\text{1-loop}}=
                   -a^2 \underbrace{\prod_{i=0}^{\kappa}\,\prod_{j=0}^{\kappa-i}}_{(i,j)\neq (q,p)}\big(a+(p-j)\ve_1+(q-i)\ve_2\big)
	           \underbrace{\prod_{i=0}^{\kappa-3}\,\prod_{j=0}^{\kappa-3-i}}_{(i,j)\neq (q-1,p-1)}-\big(a+(p-1-j)\ve_1+(q-1-i)\ve_2\big).		   
\ee
where $\kappa=p+q+r$.

It is harder to see the analytic structure of $ Z^{\mathbb{C}^2}_{\text{inst}}$ 
due to the complicated combinatorial formulation of \eqref{Z-inst}.
Great insights can be obtained by the use of AGT correspondence \cite{Alday:2009aq}
that relates the instanton partition function with conformal blocks of Virasoro algebra (see appendix).
The poles of the conformal block are in one-to-one correspondence with degenerate fields
and these latter are classified by theorem \ref{theo} in the appendix.
Eventually the poles and the associated residues for the instanton partition function
can be obtained via AGT by the Zamolodchikov's recursion relation
for Virasoro conformal blocks \cite{Zamolodchikov:1985ie, Poghossian:2009mk}
\begin{align}
\label{ZRubik}
Z_\text{inst}\big(\copl;a,\ve_1,\ve_2\big)&=
1-\sum_{m,n=1}^\infty \frac{\copl^{mn} R_{m,n}\,
Z_\text{inst}\left(\copl;m\ve_1-n\ve_2,\ve_1,\ve_2\right)}
{\big(a-m\ve_1-n\ve_2\big)\big(a+m\ve_1+n\ve_2\big)}
\\
R_{m,n}&=
2\underbrace{\prod_{i=-m+1}^m\prod_{j=-n+1}^n}_{(i,j)\neq\{(0,0),(m,n)\}}
\frac{1}{\big(i\ve_1+j\ve_2\big)}\, .
\end{align}
Using this fact, one can easily see that the product of the three copies of
$Z^{\mathbb{C}^2}_{\text{inst}}(\copl^\li\,;{a^\li},\ve^\li_1,\ve^\li_2 ) $
contributes with a triple pole in the origin when the (semi-)stability conditions are satisfied
\be\label{instanton-res}
Z^{\mathbb{P}^2}_{\text{inst}}=-\frac{1}{a^3}\,\copl^{pq+pr+qr} \,
\tilde R^{(0)}_{p,q}\, \tilde R^{(1)}_{q,r} \, \tilde R^{(2)}_{r,p} \,
Z_{\text{Res}}
					      +O\left(\frac{1}{a^2}\right)
\qquad\text{where}\qquad
\tilde R^\li_{m,n}=\frac{1}{a^{\li}+m\ve_1^\li+n\ve_2^\li}R^\li_{m,n}
\ee
and
\begin{align}
Z_{\text{Res}}&=Z_\text{inst}\big(\copl^{(0)};a^{(0)}_\text{res},\ve_1,\ve_2\big)
                Z_\text{inst}\big(\copl^{(1)};a^{(1)}_\text{res},\ve_2-\ve_1,-\ve_1\big)
	       Z_\text{inst}\big(\copl^{(2)};a^{(2)}_\text{res},-\ve_2,\ve_1-\ve_2\big),
\label{ZRES} \\[2mm]
a^{(0)}_\text{res}&=p\ve_1-q\ve_2, \qquad
a^{(1)}_\text{res}=q(\ve_2-\ve_1)+r\ve_1, \qquad
a^{(2)}_\text{res}=-r\ve_2-p(\ve_1-\ve_2). \label{ares}
\end{align}

\subsection{Results and non-equivariant limit}

\begin{figure}[!h]
\centering
\begin{tikzpicture}[scale=0.6]

\definecolor{purplex}{rgb}{0.5,0,0.7}

\draw [->] (0,-3.5) -- (0,3.5) node[right] {\large $\epsilon_2$};
\draw [->] (-3.5,0) -- (3.5,0) node[above] {\large $\epsilon_1$};


\fill[black!40!white,opacity=0.7] (0,1.25) -- (1.75,-.5) -- (1.75,-1) -- (-.5,-1) -- (-1.5,0) -- (-1.5,1.25) --(0,1.25);

\node at (-.2,0) {\small $U$};


\draw[dashed] (-3.5,1.25) node[left] {$r_4$} -- (3.5,1.25);
\draw[dashed] (-3.5,-1) node[left] {$r_3$} -- (4,-1);

\draw[dashed] (1.75,-3.75) node[below] {$r_2$} -- (1.75,3.5);
\draw[dashed] (-1.5,-3.75) node[below] {$r_1$} -- (-1.5,3.5);

\draw[dashed] (-3.5,2) -- (2.25,-3.75)  node[right] {$r_6$};
\draw[dashed] (-2.25,3.5) -- (3.5,-2.25) node[right] {$r_5$};


\draw[thick] (1.75,1.25) -- (1.75,-3.25) -- (-2.75,1.25) -- (1.75,1.25);

\draw[thick] (-1.5,-1) -- (2.25,-1) -- (-1.5,2.75) -- (-1.5,-1);

\end{tikzpicture}
\caption{The region $U$ is the intersection of two rectangle triangles,
one is defined by the three lines $r_1=\{x=-p+1\}$, $r_3=\{y=-q+1\}$ and $r_5=\{y=-x+r-1\}$
and the other by the three lines $r_2=\{x=p\}$, $r_4=\{y=q\}$ and $r_6=\{y=-x-r\}$.}\label{Uset}
\end{figure}
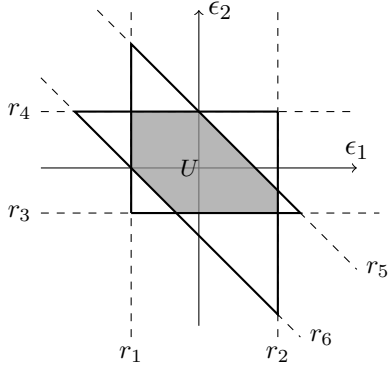

From the discussion on the previous subsection, the integrand of \eqref{Zfull} exhibits a single pole in the origin.
The integral is then easily computed via residue evaluation making use of \eqref{ZRubik}
and the final result in the case of odd first Chern class is
\be\label{result-P2-DI}
\begin{aligned}
Z^{\mathbb{P}^2}_{\mathcal{N}=2}(\copl,x,z,\ve_1,\ve_2)\big|_{c_1=1}
=&\copl^{-\frac{1}{4}c_1}\sum_{\substack{p+q+r+c_1=\text{even} \\ \{p, q, r|\text{stable}\}}}
        \copl^{-\frac{1}{4}(p^2+q^2+r^2-2pq-2pr-2qr)} \\
 &\exp\Bigg(\!-\frac{1}{4}\sum_{\ell=0}^2
               \frac{(a_\text{res}^\li)^2\,\imath^*_{P_{(\ell)}}(\alpha z+p x)}{\ve_1^\li \ve_2^\li}\Bigg)
     \Bigg(\prod_{(i,j)\in(U\cap\mathbb{Z}^2)\setminus(0,0)} \frac{1}{i\ve_1+j\ve_2}\Bigg)
    Z_{\text{Res}}
\end{aligned}
\ee
where $Z_{\text{Res}},a_\text{res}^\li$ are defined in \eqref{ZRES},\eqref{ares}
and the subset $U$ of $\mathbb{R}^2$ is depicted in figure \ref{Uset}.

Formula \eqref{result-P2-DI} agrees with the generating function of equivariant Donaldson invariant with $c_1=1$
on $\mathbb{P}^2$ computed in \cite{Gottsche:2006tn}.
In the non-equivariant limit (i.e. $\ve_1,\ve_2\to 0$)
it reproduces correctly
ordinary Donaldson invariants for $SO(3)$-bundle on $\mathbb{P}^2$
computed in \cite{1995alg.geom..6019E}


The calculation can be repeated also in the case of even first Chern class.
This case is subtle because it allows for reducible connections and singularities in the moduli space.
The result of the integral \eqref{Zfull} in this case has two contributions: 
the first one is the contribution from strictly stable bundles
($p,q,r$ satisfying strict triangular inequalities in \eqref{stability-p2}), it has  
the same expression \eqref{result-P2-DI} but with $c_1=0$.
The second term is the contribution of strictly semistable bundles
that are reducible connections
($p,q,r$ saturating one of the inequalities in \eqref{stability-p2}),
it is given by
\be\label{result-P2-DI-ss}
\begin{aligned}
&\frac{1}{2}\!\!\!\!
\sum_{\substack{p+q+r=\text{even} \\ \{p, q, r|\text{strictly semistable}\}}}
        \copl^{-\frac{1}{4}(p^2+q^2+r^2-2pq-2pr-2qr)}
\exp\Bigg(\!-\frac{1}{4}\sum_{\ell=0}^2
               \frac{(a_\text{res}^\li)^2\,\imath^*_{P_{(\ell)}}(\alpha z+p x)}
                    {\ve_1^\li \ve_2^\li} \Bigg) \\
&\big(\text{sgn}(p-q-r+1)p\epsilon_1+\text{sgn}(q-p-r+1)q\epsilon_2\big)
\Bigg(\prod_{(i,j)\in(U\cap\mathbb{Z}^2)\setminus(0,0)} \frac{1}{i\ve_1+j\ve_2}\Bigg)
    Z_{\text{Res}}
\end{aligned}
\ee
where $\text{sgn}$ is the sign function and
$Z_{\text{Res}}$, $a_\text{res}^\li$ and $U\subset\mathbb{R}^2$ are the same as in \eqref{result-P2-DI}.
The factor $\frac{1}{2}$ in front of the sum
is necessary because reducible connections of $U(2)$-bundle are counted twice
in the path integral \cite{Bershtein:2015xfa}
and so we need to normalize their contribution.

This formula is conjectured to be the generating function of equivariant
Donaldson invariants with $c_1=0$ on $\mathbb{P}^2$ and appears new in the literature.
In the non-equivariant limit it correctly reproduces ordinary Donaldson invariants
for $SU(2)$-bundle on $\mathbb{P}^2$, that are calculated in \cite{1995alg.geom..6019E}
\be\label{limitSU2}
\begin{aligned}
\lim_{\ve_1,\ve_2\to 0}\hspace{-4mm}&\hspace{4mm}
 Z^{\mathbb{P}^2}_{\text{full}}(\copl,x,z,\ve_1,\ve_2)\big|_{c_1=0}= \\
=&\copl\left(-\frac{3}{2}z\right)
   +\copl^2\left(-\frac{13}{8}\frac{x^2 z}{2!}-\frac{x z^3}{3!}+\frac{z^5}{5!}\right) \\
+&\copl^3\left(-\frac{879}{256} \frac{x^4 z}{4!} - \frac{141}{64} \frac{x^3 z^3}{3!\, 3!} - \frac{11}{16} \frac{x^2 z^5}{2!\, 5!} 
                 +\frac{15}{4} \frac{x z^7}{7!} + 3 \frac{z^9}{9!}\right) \\
+&\copl^4\left(-\frac{36675}{4096}\frac{x^6 z}{6!}-\frac{1515}{256}\frac{x^5 z^3}{5!\, 3!}
                 -\frac{459}{128}\frac{x^4 z^5}{4!\, 5!}+\frac{51}{16}\frac{x^3 z^7}{3!\, 7!}
                 +\frac{159}{8}\frac{x^2 z^9}{2!\, 9!}+24\frac{x z^{11}}{11!}+54\frac{z^{13}}{13!}\right)
+O(\copl^5).
\end{aligned}
\ee

\section{Preliminary results for $U(2)$ $\N=2$ theory on $\mathbb{F}_0$}

We tried to repeat the computation of the previous section for other compact toric surfaces.
Here we present some partial results obtained for $\mathbb{F}_0=\mathbb{P}^1\times\mathbb{P}^1$.

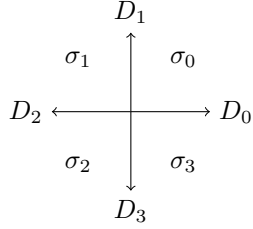
\begin{figure}[!h]
\centering
\begin{tikzpicture}[scale=0.7]
\draw [->] (0,0) -- (1.5,0) node[right] {$D_0$};
\draw [->] (0,0) -- (0,1.5) node[above] {$D_1$};
\draw [->] (0,0) -- (-1.5,0) node[left] {$D_2$};
\draw [->] (0,0) -- (0,-1.5) node[below] {$D_3$};

\node at (1,1) {$\sigma_0$};
\node at (-1,1) {$\sigma_1$};
\node at (-1,-1) {$\sigma_2$};
\node at (1,-1) {$\sigma_3$};
\end{tikzpicture}
\caption{Toric fan of $\mathbb{F}_0$.
$\sigma_\ell$ labels the cone of dimension two relative to the $\ell$-th $\mathbb{C}^2$ coordinates patch.}
\label{figure4}
\end{figure}

The homogeneous coordinates are denoted by $([z_0:z_1],[w_0:w_1])$.
The full $U(2)$ partition function is given by 
the contour integral \eqref{Zfull-gen}
where the integrand is a product of four copies of the Nekrasov partition function
\be\label{Z-full-F0}
Z_\text{full}^{\mathbb{C}^2}\big(\copl^\li\,;{a}^\li,\ve_1^\li,\ve_2^\li\big)
\ee
where $\ve_1^\li,\ve_2^\li$ are the generators of $(\mathbb{C}^\ast)^2$
action on $T_{P_\li}\mathbb{F}_0$,
which can be read from the fan (Fig.~\ref{figure4}).
\be\label{weights-F0}
([z_0:z_1],[w_0:w_1])\to([z_0:e^{\ve_1}z_1],[w_0:e^{\ve_2}w_1]) \quad \iff \quad
\begin{array}{l|r|r|r|r}
\ell        &   0   &    1   &    2   &   3 \\ \hline 
\ve_1^\li   & \ve_1 & -\ve_2 & -\ve_1 &  \ve_2 \\
\ve_2^\li   & \ve_2 &  \ve_1 & -\ve_2 & -\ve_1 \\ \hline
\end{array}
\ee
and $a_\alpha^\li$ are the generators of the $(\mathbb{C}^\ast)^2$ framing action
\be
\begin{aligned}
\vec a^{(0)}&=\vec a+\vec k^{(1)}\ve_1+\vec k^{(2)} \ve_2, &\quad
\vec a^{(1)}&=\vec a-\vec k^{(2)} \ve_2+\vec k^{(3)}  \ve_1, \\
\vec a^{(2)}&=\vec a-\vec k^{(3)} \ve_1-\vec k^{(4)} \ve_2, &\quad
\vec a^{(3)}&=\vec a+\vec k^{(4)} \ve_2-\vec k^{(1)} \ve_1,
\end{aligned}
\ee
where $\vec k^\li$ are the magnetic fluxes.
The coupling in \eqref{Z-full-F0} $\copl^\li=\copl\, e^{\imath^*_{P_{(\ell)}}(\alpha_1 z_1 +\alpha_2 z_2 + p x)}$
is shifted by the insertion of the observables described in section \ref{section-obs}
evaluated at the $(\mathbb{C}^\ast)^2$-fixed points $P_{(\ell)}$ of $\mathbb{F}_0$.

In principle one can repeat the same steps of section \ref{section-p2}
and compute the residue at the origin of the full partition function
using Zamolodchikov's recursion relation \eqref{ZRubik}.


This is possible as long as only simple poles appear at the origin.
Indeed the difference with respect to the case of $\mathbb{P}^2$ is that 
now the integrand in \eqref{Zfull-gen} has, in general, a higher order pole at the origin.
But for certain choices of the polarization $H$ the first orders in the instanton expansion
present only simple poles and some partial result can be obtained.

For example, working with the polarization $H= D_0+D_1$ and the first Chern class $c_1=D_1$,
higher order poles appear only from instanton number three.
We are therefore able to obtain the expressions for equivariant Donaldson invariants
only up to two instantons
and the result agrees with the one computed making use of the wall-crossing formula of \cite{Gottsche:2006tn}.
In the non-equivariant limit it reproduces correctly the first few ordinary Donaldson invariants
\begin{equation}
\label{F0-DI}
\begin{aligned}
&\lim_{\epsilon_1,\epsilon_2\to 0}
\text{Res}\Big(Z^{\mathbb{F}_0}_\text{full}\big(a,\epsilon_1,\epsilon_2;
\copl e^{\alpha_1 z_1 +\alpha_2 z_2 +p x}\big)\Big|a=0\Big)=\\
&=\frac{\copl}{1!}\left(\frac{z_1}{1!}
-\frac{1}{2}\frac{z_2}{1!} \right) \\
&+\frac{\copl^2}{2!}\left[
\frac{x^2}{2!}\left(\frac{z_1}{1!}\frac{9}{4}-
\frac{z_2}{1!}\frac{17}{16}\right)
+\frac{z_1^2}{2!}\frac{z_2^3}{3!}1
-\frac{z_1}{1!}\frac{z_2^4}{4!}\frac{7}{4}
+\frac{x}{1!}\left(\frac{z_1^2}{2!}\frac{z_2}{1!}1-
\frac{z_1}{1!}\frac{z_2^2}{2!}\frac{1}{4}-
\frac{z_2^3}{3!}\frac{7}{16}\right)
+\frac{z_2^5}{5!}\frac{31}{16}
\right]
\\
&+O(\copl^3) \, .
\end{aligned}
\end{equation}

To compute higher orders in \eqref{F0-DI} we need to understand properly
how to include the residues of higher order poles in \eqref{Zfull-gen}.
The problem is that for these contributions it is not clear how to associate
correctly stability conditions.
Further investigations on this point are necessary.

\section{$\N=2^\star$ theory, Euler characteristic and mock modular forms}

The $\N=2^\star$ theory is obtained adding to the pure Super--Yang--Mills theory
a hypermultiplet of mass $M$ in the adjoint representation.
In the massless limit the theory becomes a $\N=4$ Super--Yang--Mills
whose partition function is known to be the
generating function of the Euler characteristic of the moduli space of unframed semi-stable
equivariant torsion-free sheaves~\cite{Vafa:1994tf}.

In \cite{Bershtein:2015xfa} we reproduce the computations of section \ref{section-p2}
on the complex projective plane for the $\N=2^\star$ theory.
Again the full partition function is given by a contour integral where the integrand is
made by copies of the Nekrasov partition function for a theory with a hypermultiplet in the adjoint
representation of the gauge group.

To read the analytic structure of the instanton partition function
we make use again of the AGT correspondence
and of the Zamolodchikov's recursion relation for conformal blocks on the one-punctured torus \cite{Zamolodchikov:1985ie}.
The latter can be written \cite{Poghossian:2009mk}
\be\label{ZRubik-adj}
Z^{\mathbb{C}^2}_{\text{inst,adj}}(\copl;\, a,M,\epsilon_1,\epsilon_2)=
\big(\copl^{-1/24}\eta(\copl)\big)^{-2\frac{(M-\ve_1)(M-\ve_2)}{\ve_1\ve_2}}
H(\copl;\, a,M,\ve_1,\ve_2),
\ee
where $\eta(q)=q^{1/24}\prod_{n=1}^\infty(1-q^n)$ is the Dedekind eta function
and
\be
H(\copl;\, a,M,\ve_1,\ve_2)=1-\sum_{m,n=1}^\infty \frac{\copl^{mn} R^\text{adj}_{m,n}\,
                           H\left(\copl;\, m\ve_1-n\ve_2,M,\ve_1,\ve_2\right)}
                           {\big(a-m\ve_1-n\ve_2\big)\big(a+m\ve_1+n\ve_2\big)}
\ee
with
\be\label{Rfactor-adj}
R^\text{adj}_{m,n}=2\Bigg(\prod_{i=-m+1}^m\prod_{j=-n+1}^n\big(M-i\ve_1-j\ve_2\big)\Bigg)/
\Bigg(\underbrace{\prod_{i=-m+1}^m\prod_{j=-n+1}^n}_{(i,j)\neq\{(0,0),(m,n)\}}
 \big(i\ve_1+j\ve_2\big)\Bigg).
\ee

Repeating what done in section \ref{section-p2},
evaluating the residue at the origin,
taking the massless limit $M\to 0$
and since
\be
\lim_{M\to 0}H\left(\copl;\, m\ve_1-n\ve_2,M,\ve_1,\ve_2\right)=1
\ee
we get
\be\label{Result-N2star}
\lim_{M\to 0}\frac{1}{M}\text{Res}\big(Z^{\mathcal{N}=2^\star}_\text{full}(\copl;\, a,M,\ve_1,\ve_2)\big|a=0\big)
=\big(\copl^{-1/24}\eta(\copl)\big)^{-6} \copl^{-\frac{1}{4}c_1^2}\copl^{-\frac{1}{4}(p^2+q^2+r^2-2pq-2pr-2qr)}.
\ee

The $\N=4$ partition function on the complex projective plane
is obtained summing the result \eqref{Result-N2star} over the fluxes $p,q,r$
satisfying (semi-)stability conditions \eqref{stability-p2}.
The result is
\be\label{n=4}
Z_{\mathcal{N}=4}^{\mathbb{P}^2}(\copl) = \big(\copl^{-1/24}\eta(\copl)\big)^{-6} 
\sum_{c_1=0,1}\Bigg(\sum_{\substack{\{p, q, r\}\\ \text{strictly stable}}}+
              \hspace{7mm}\frac{1}{2}\hspace{-7mm}
              \sum_{\substack{\{p, q, r\}\\ \text{strictly semi-stable}}}\Bigg)
\copl^{-\frac{1}{4}(1-2c)c_1^2}\copl^{-\frac{1}{4}(p^2+q^2+r^2-2pq-2pr-2qr)}
\ee
where
--- like in \eqref{result-P2-DI-ss} ---
the factor $\frac{1}{2}$ in front of the sum on strictly semi-stable bundles
is necessary because reducible connections of $U(2)$-bundle are counted twice
in the path integral \cite{Bershtein:2015xfa}.

Formula \eqref{n=4} can be written in terms of Hurwitz class number $H(D)$
and it reproduces the results obtained in \cite{Yoshioka1994,Yoshioka1995}
via finite fields methods
\begin{align}
&Z_0(\copl)=\big(\copl^{-1/24}\eta(\copl)\big)^{-6}\sum_{n=0}^\infty 3 H(4n)\copl^n      && c_1=0,         \\[2mm]
&Z_1(\copl)=\big(\copl^{-1/24}\eta(\copl)\big)^{-6}\sum_{n=0}^\infty 3 H(4n-1)\copl^n    && c_1=1.
\end{align}
These formulas were studied in \cite{Vafa:1994tf}, they are mock modular holomorphic form,
namely Eisenstein series of weight 3/2, see  \cite{Bringmann:2010sd}.

\section*{Acknowledgments}
A.T. thanks the organizers and participants of the conference ``Interactions between Geometry and Physics'', in honor of Ugo Bruzzo's 60th birthday,
17-22 August 2015, Guaruj\'a, S\~{a}o Paulo, Brazil, for interesting discussions in Mathematics and Physics in a pleasant atmosphere, perfectly in line with Ugo's style.
M.B.~is a Young Russian Mathematics award winner and would like to thank its sponsors and jury.
M.B.~is supported by the Russian Science Foundation under the grant 14-12-01383,
G.B.~is supported by the INFN National project ST\&FI, M.R.~and A.T.~are supported by the INFN National project GAST.
M.R.~is supported by National Group of Mathematical Physics (GNFM-INdAM).

\appendix

\section{Virasoro algebra and AGT correspondence}

Virasoro algebra is defined by
\be
[L_n,L_m]=(n-m)L_{n+m}+\frac{c}{12}(n^3-n)\delta_{n+m,0}.
\ee
The representations of the Virasoro algebra are constructed from highest weights
\be
L_0|\phi_\Delta\rangle=\Delta|\phi_\Delta\rangle, \qquad
L_k|\phi_\Delta\rangle=0 \quad k>0.
\ee
The Verma module is freely generated by
\be
L_{-k_1}L_{-k_2}\cdots L_{-k_n}|\phi_\Delta\rangle, \qquad
1\le k_1\le k_2\le \dots\le k_n
\ee
where the descendants are at level $k=\sum_i k_i$.
Among these descendants there can be null vectors $|\chi_N\rangle$
at level $N$, they satisfy
\be
L_0|\chi_N\rangle=(\Delta+N)|\chi_N\rangle, \qquad
L_k|\chi_N\rangle=0 \quad k>0.
\ee

Null vectors are classified by the following theorem
\begin{thm}\label{theo}
(Kac, Feigin--Fuchs \cite[Corollary 5.2]{iohara2013representation})
Non-zero null vectors exist in the Verma module of degenerates fields
$|\phi_{\Delta_{mn}}\rangle$
\be
\Delta_{mn}=\alpha_{nm}(Q-\alpha_{nm}), \quad
\alpha_{nm}=\frac{1-n}{2}b+\frac{1-m}{2}b^{-1}, \quad
c=1+6Q^2, \quad
Q=b+b^{-1},
\ee
these are at level $N=nm$ and all non-zero null vectors are obtained in this way.
\end{thm}

Since the poles of the Virasoro conformal block are known to be in one-to-one correspondence
with degenerate fields, one can in turn understand the analytic structure
of the instanton partition function \eqref{Z-inst} using AGT relation
\be
\ve_1=b, \quad
\ve_2=b^{-1}, \quad
Q=\ve_1+\ve_2, \quad
\alpha=\frac{Q}{2}-a.
\ee

\bibliography{REf}
\bibliographystyle{JHEP}


\end{document}